# The Measurement of Italian Universities' Research Productivity by a Non Parametric-Bibliometric Methodology[1] [*]

*Giovanni Abramo[a,b,**], Ciriaco Andrea D'Angelo[a] and Fabio Pugini[a]*

[a] Technology Transfer and Entrepreneurship Lab, Dep. of Engineering and Management, University of Rome "Tor Vergata" - Italy

[b] Italian National Research Council

**Abstract**

This paper presents a methodology for measuring the technical efficiency of research activities. It is based on the application of *data envelopment analysis* to bibliometric data on the Italian university system. For that purpose, different input values (research personnel by level and extra funding) and output values (quantity, quality and level of contribution to actual scientific publications) are considered. Our study aims at overcoming some of the limitations connected to the methodologies that have so far been proposed in the literature, in particular by surveying the scientific production of universities by authors' name.

[1] Abramo, G., D'Angelo, C.A., Pugini, F. (2008). The measurement of Italian universities' research productivity by a non parametric-bibliometric methodology. *Scientometrics*, 76(2), 225-244. DOI: 10.1007/s11192-007-1942-2

[*] Authors are deeply indebted to Cristiano Giuffrida and Ilaria Croce for thier invaluable support in data extraction, codification and elaboration.

[**] Corresponding author, Dipartimento di Ingegneria dell'Impresa, Università degli Studi di Roma "Tor Vergata", Via del Politecnico 1, 00133 Rome - ITALY, tel. +39 6 72597362, abramo@disp.uniroma2.it

# 1. Introduction

The process of evaluation of scientific research has become a central element in the management and governance policies of national research systems and, consequently, of individual institutions, individual disciplinary areas within each institution and individual organization units (faculties, departments, etc.) within each area. This is especially true for publicly funded activities. The guidelines and reforms involving the funding systems, the transformation processes within the organizations in charge of advancing scientific frontiers, and finally the more and more stringent requirement of evidence for the socio-economic impact of publicly funded activities, all support a genuine "demand for assessment" which has arisen in all developed countries, albeit with different connotations and characteristics.

The most widespread evaluation methodologies can be classified into two general types: the one known as the bibliometric methodology, and the peer-review methodology. Both have pros and cons, extensively discussed in the literature (Horrobin, 1990; Moxham and Anderson, 1992; MacRoberts and MacRoberts, 1996; Moed, 2002; Van Raan, 2005), in terms of costs, execution times, limitations and objectiveness of measurement. At international level, peer-review type initiatives for the assessment of scientific productivity in academic systems have been reported, such as those undertaken in the UK starting from 2001, the Research Assessment Exercise, RAE (www.rae.ac.uk); or mixed-type initiatives, in which peer-review methodologies were supported by the use of bibliometric indicators, as in the case of the study on Dutch chemistry and chemical engineering laboratories performed by the Leiden University (Vsnu, 2002). Pure bibliometric techniques have been adopted, among others, in work by Abbott and Doucougalios (2003) and Worthington and Lee (2005) for the Australian university system, by Flegg et al. (2004) and Athanassopoulos and Shale (1997) for the UK, and Baek (2006) for the US. In these studies, scientific productivity is assessed by means of total factor productivity indexes, calculated with DEA (data envelopment analysis) non-parametric techniques.

Within the Italian public research system, and the university system in particular, the process of development and application of scientific production assessing techniques is still far from reaching effective, robust, low-cost and universally agreed-upon results.

The outcomes of the first large-scale evaluation process were published in 2006[2]. That exercise was performed by the Steering Committee for Research Evaluation (CIVR)[3], an agency controlled by the Italian Ministry for Research and University Education, in charge of promoting research evaluation activities and support to quality and more effective utilization of national research. Based on the model of British RAE, a peer-review approach was adopted by CIVR, which was able to draw upon human and financial resources as needed. There is nonetheless still limited consensus on the appropriateness of the applied methodology, on the effectiveness of the indicators, on the significance of the results, and finally on the possible advantages in its direct application in resource allocation (Abramo, 2006).

On the other hand, bibliometric techniques, simpler and remarkably less cost-intensive, were used in other studies on scientific productivity, such as the assessment exercise by the Association of the Deans of the Italian Universities (Crui), in which non-normalized productivity indexes (pro-capita publications) and impact indexes (pro-capita citations) of national universities were calculated (Crui, 2002) on the basis of aggregated extractions by disciplinary area from Thomson Scientific's Science Citation Index (SCI) database. Abramo and Pugini (2004) integrated the assessment of scientific productivity with the analysis of technological productivity (measured in terms of number of patents filed): in the study, an aggregate approach at university level was used, and single output/single input non-normalized productivity indicators were calculated. Bonaccorsi et al. (2006), on the other hand, applied DEA-type non-parametric techniques for the assessment of total factor productivity indicators to the Italian National Research Council (Cnr)[4] and the university system. In that case, productivity indicators were calculated directly at aggregate university level: the aim was not so much to perform a comparative analysis of scientific productivity, as to evaluate possible economies of scale, and the trade-off between research activities and teaching within universities.

The objective of this study is to develop a methodology for measuring the performance of research activities in the public sector, which we will call a bibliometric-non-parametric methodology; to apply it to the Italian university system, while overcoming the limitations connected to the methodologies used so far; to conduct a number of sensitivity tests, by altering independent variables. This methodology, as all bibliometric ones, has a host of advantages over a peer-review approach: it is low-cost, non-invasive, easy to implement, ensures rapid updates and time-series comparisons, is based on objective qualitative-quantitative data, has a high degree of representativeness of the surveyed universe, allows of international comparisons. On the other hand, as it is based on international scientific publications alone, it does not include other types of research "products".

---

[2] http://vtr2006.cineca.it/
[3] www.civr.it
[4] Cnr is the main public research institution in Italy, staffing around 8,000 employees in its over 100 research institutes, carrying out research in all main fields of science.

In the context of bibliometric techniques, the innovative contribution of this work to the international state of the art is twofold. First, the assessment of the scientific efficiency of universities is developed in two different steps: upstream, by applying the DEA methodology to individual university disciplinary areas; downstream, by constructing a global efficiency index by university as an appropriate weighted combination of the scores obtained in individual areas. This has made it possible to overcome technical and methodological limitations connected with the heterogeneity in the scientific composition of the various universities, the varying degrees of publication prolificity among different areas and the representativeness by area of the journals surveyed in the source databases, thus enabling a more consistent and robust comparative analysis of the universities. Secondly, in a completely original approach with respect to homologous large-scale exercises, a "bottom-up" approach was adopted for the definition of input and output data to be used in the model (whose importance is crucial for the accuracy of DEA bibliometric assessments): the publications were thus directly associated with the relevant authors, and bibliometric values were aggregated by disciplinary area and by university only later, on the basis of the affiliation of each single author[5]. Such process led to the generation of a publication archive based on authors' names rather than disciplinary areas. This made it possible, when confronting research efficiency of universities in each disciplinary area, to measure all papers authored by personnel falling in that specific area, rather than the publications falling in it. This methodology thus overcomes the limitations inherent in previous studies where the aggregate productivity of an area was measured on the basis of portfolios assessed counting the publications falling in that area[6] (Crui, 2002; Abramo and Pugini, 2004; Bonaccorsi et al., 2006). The database thus assembled represents a major breakthrough in the scientific landscape of this sector, and ensures high confidence on the significance of the input and output measures used in the model.

---

[5] In Italy, universities' research personnel is classified according to the scientific sector of her/his speculation activities.

The remaining part of this work is articulated as follows. Section 2 surveys the two main methodologies for the assessment of research activities, and discusses their strengths and weaknesses. Section 3 is devoted to a more in-depth discussion of parametric and non-parametric techniques for the assessment of total productivity, with particular emphasis on DEA model. Section 4 illustrates the domain of investigation, the approach used to process output data for the creation of a researcher-based publication database and the DEA model. Results of the analysis are presented in Section 5, whereas observations and conclusions by the authors are presented in Section 6.

## 2. Methods of assessment of research activities

### 2.1 Peer-review methodologies

Peer-review methodologies are based on the assessment of research outputs of research organizations, by panels of assessors selected by the authority presiding the assessment.

As such methodologies depend on the quality judgment of experts, they might suffer severe limitations (Moxham and Anderson, 1992; Horrobin, 1990), most of which can be traced back to the subjectivity constraint on that judgment. Such subjectivity operates at three levels: upon selection of the experts who will assess each product; upon assessment of the level of excellence of those products, performed by the peers; during the preceding process of selection of the products to be submitted to assessment, performed by each individual research unit. Subjective assessment might be affected by actual or potential conflicts of interests, by the tendency to rate products by famous and renowned scholars higher than those by younger, lesser-known researchers, or by the failure to acknowledge particular qualitative aspects of the product (which become the more important the more specialized the work is). The methodology, furthermore, presents no universality, as the mechanisms of appraisal assignment are defined independently by the assessing panel, and are therefore open to possible distortions. The times and costs for this type of assessment exercises represent another critical element in the methodology.

---

[6] The method of aggregation of publications by disciplinary area based on the scientific category associated to a particular journal, while more straightforward and convenient, induces significant distortions (tested by the authors of this study) in productivity measures; this is due to the fact that researchers fairly frequently publish papers falling in disciplinary areas other than those they belong to.

**2.2 Methodologies based on bibliometric techniques**

Research scientists, especially those affiliated to public laboratories, usually disseminate the results of their projects via publication in scientific journals, preferably international and prestigious ones. This is, in short, the basic assumption of bibliometric approaches to research assessment. Such approaches utilize more or less sophisticated qualitative-quantitative indicators, linked to two basic drivers: the publication in itself, and any citations obtained over time. The survey of publications and citations is usually performed by querying ad hoc databases, such as those built by Thomson Scientific[7].

Technical and methodological limits embedded in bibliometric indicators have been amply analised (Van Raan, 2005). The most crucial limitation, which consequently has the largest impact in the event of use for assessment/allocation purposes, is to consider the scientific publication alone as a proxy of overall research output, while overlooking all other forms of dissemination, both codified (essays, scientific texts, electronic publications, technical reports, protocol, databanks, patents etc.) and non-codified (consulting, training, technical services, etc.) This hypothesis has nonetheless ample empirical evidence to warrant its assumption. On the one hand, publication is by far the most common means used by researchers in Europe and the United States, where the other measurable forms of codification (such as patents) contribute less than 10% to the total transfer of new scientific and technological knowledge (Cohen et al., 2002;

---

[7] The main ones are: the Science Citation Index (SCI) for technical and scientific disciplines, the Social Science Citation Index and the Art and Humanities Citation Index for social sciences and humanities; http://www.isinet.com/

Agrawal ed Henderson, 2002; Colyvas et al., 2002) from universities to companies. On the other hand, it has been shown that, in countries with considerable scientific-technological production, there is a significant correlation between patent production and scientific production. In other words, universities with the highest publication intensity also have the highest patenting intensity (Adams and Griliches, 1998; Lach and Shankerman, 2003). As concerns Italy, the situation is partly different. The recourse to forms of protected codification of public research results, especially in universities, is negligible.[8] Furthermore, publication intensity tends to increase more strongly in Italy than in other countries.[9] It can thus be concluded that scientific publications always represent, especially in Italy, a robust indicator of the production of new knowledge developed within the public research system.

A correct measurement of publications by means of databases such as the SCI is, moreover, subject to limitations which are intrinsic to the reference database: in the SCI about 4,800 international journals are covered, which cannot be considered an exhaustive sample of the complex scientific publication universe; the representativeness of the journals covered varies according to discipline, and is definitely higher in technical-scientific areas than in the humanities, where it is rather marginal.

These methodological limitations are compounded by technical constraints affecting the accuracy of bibliometric surveying. These usually depend on the wrong attribution of publications and attendant citations to the research organizations their authors belong to: in particularly critical situations, the percentage of attribution errors can reach 30% (Moed, 2002). Such errors are potentially determined by joint causes, including but not limited to: i) wrong database entry of information specific to the publication (affiliation institutions, authors' personal data, citations, etc.); ii) extreme variance in the indication of the affiliation institution (full or abbreviated names in the original language or in English), or, conversely, identical indications for different institutions, for example universities located in the same city; iii) indication of the affiliation to research centers which naturally fall within a larger institution (for example, laboratories, departments, institutes, medical centers, etc.); iv) wrong indication of affiliation by the author (for instance, this happens when researchers assigned to a university, but in fact employed by a different research institution, state that they are affiliated to the university, and vice versa).

---

[8] In 2001, the number of patents produced by the whole of Italian universities equaled that of the University of Wisconsin; the total figure of universities and public research institutions ranked lower than that of the Massachusetts Institute of Technology (MIT) alone, whose research budget approximately matches that of CNR (Abramo and Pugini, 2005).

[9] Scientific publications by American universities decreased by 10% in the 1995-2000 period, as against an increase in research expenditure by 22% during the same period, at constant prices. In the same period, scientific publications decreased by 9% in Canada, by 5% in the Netherlands and by 1% in the UK (Oecd, 2003; Nsb, 2004). Italy, on the other hand, showed the highest annual publication growth rate among G7 countries during the same period, coming in second to the UK in the number of publications per researcher.

Previous bibliometric studies on the scientific production of Italian universities (Crui, 2002; Abramo and Pugini, 2004; Bonaccorsi and Daraio, 2003) were based on extraction from the SCI database and on a data post-coding process involving a top-down approach, in which the attribution of a publication to a research organization depended on the identification of its name in the "address" specified by its authors. Such approach can solve only part of the set of problems involved in the process of accurately attributing a publication to its authors, which is collectively known as "*disambiguation*" in the literature (Torvik et al., 2005).

Finally, even if the above technical limitations are mitigated/reduced, the methodological limitations shortly mentioned above still remain in place. In particular, a consequence of the varying degrees of prolificity among scientific disciplines and variegated representation of different areas in terms of journals covered is the following: if the bibliometric measurements are made at aggregate level, as in the case of university-level assessments, the subsequent ranking might be significantly distorted by the different distribution of (human and financial) resources among the various scientific areas of each university.

## 3. Parametric and non-parametric techniques for the assessment of total productivity

Within the realm of bibliometric techniques, the most immediate approach for the measurement of the scientific productivity of research organizations involves the calculation of partial indexes, through the normalization of output data for particular input values, such as R&S expenditure and/or the number of research staff. The variability among the compared elements also suggests using more articulated and robust methodological approaches (Gauffriau and Larsen, 2005). Research institutions in general, and universities in particular, are complex realities, affected by a number of different inputs and outputs. The total productivity of production factors is therefore not easily measured through the construction of weighted indexes based on partial measures (Bonaccorsi, 2003; Bonaccorsi et al., 2006). There are two traditional approaches used by econometrists to measure the total productivity of research: parametric and non-parametric techniques.

Parametric methodologies are based on the a priori definition of a function representing the relationship between input and output of a particular production unit most effectively. These estimation processes have the purpose of determining the coefficient (model parameters) of a regression equation describing the production function, usually a Cobb-Douglas type equation. The main limitation of such methodology lies in the need to define in advance closed models describing the production function: this entails the need to make assumptions on the relationship between input and output, for instance to assume additive inputs rather than a linear function connecting the two values. Furthermore, parametric techniques cannot identify benchmark best practices, but define expected (or optimal) performances at selected input levels.

The purpose of non-parametric methods, on the other hand, is to compare empirically measured performances of production units (commonly known as Decision Making Units, DMUs), in order to define an "efficient" production frontier, comprising the most productive DMUs. The reconstruction of that frontier is useful to assess the inefficiency of the other DMUs, based on minimum distance from the frontier. The main advantages of non-parametric methods can be summarized as follows:

- Complex production systems with multiple inputs and outputs are assessed by means of a single global efficiency value, the Total Factor Productivity, obtained with no pre-defined weighting factors of any sort.
- No functional relationship needs to be established to define production processes, nor do optimization or estimation processes.
- The frontier from which efficiency coefficients are calculated is obtained from actually measured DMUs; in other words, a comparison is made between real production units that can be used as references for best practices.

At the same time, correct identification of inputs and output indicators is crucial to the reliability of the model application.

One of the non-parametric methods most commonly observed in the literature is the DEA. Developed as a technique for assessing the efficiency of industrial production systems (Charnes et al., 1978; Banker et al., 1984), the DEA has extremely limited applicability hypotheses: i) homogeneity of DMUs: the production units must produce the same type of goods or services using the same type of resources; ii) convexity of the analyzed set: the frontier includes all possible linear combinations of the efficient units; iii) free disposability, or the possibility to eliminate resources with no costs.

In this study, the output-oriented DEA model has been applied. In that model, the efficiency deviation from the frontier is evaluated as the maximum equiproportional increase of all outputs as allowed by the available inputs. This model is particularly appropriate for scientific research, since the overall objective is not to reduce the input while maintaining constant production, but to maximize production with the resources available. The DEA methodology includes two distinct models for cases of absence (CRS) or presence of returns to scale of production factors (VRS). The use of the CRS specification when not all DMUs are operating at optimal scale, will result in measures of *technical efficiency* (TE) which are confounded by *scale efficiency* (SE). The use of the VRS specification will permit the calculation of TE devoid of these SE effects. The SE can be extracted by applying both models to the same data set. The problem of calculating the frontier and the DEA efficiency indexes can be formulated in terms of linear programming and is easily solved by using specially developed software. In particular, for our analysis, the Efficiency Measurement System (EMS) developed by the University of Dortmund was utilized (Scheel, 2000). The use of the DEA method should, in any case, be supported by technical-methodological comments which can help correctly interpret any results arising out of it. First, the DEA is of purely deterministic nature: any deviation from the frontier is associated with inefficiency, and it is not possible to take into consideration casual elements or external noise which might have affected the results. Secondly, the calculated efficiency measure is only valid for the variables that are measured and used by the model. While representing measures of total productivity, those values depend exclusively from the choice of variables, and might therefore not give a completely representative picture of the efficiency of DMUs, especially as important input or output factors could be overlooked. In the specific case of the bibliometric-type measurement of the production performance of Universities with the DEA model, possible distortions might, for instance, arise if: (on the input side) time is allocated incongruously between research and teaching or between different types of research (basic/applied), or production factors overlooked in the model are non-homogenously available, such as scientific instruments, or non-employed staff (PhD students, external collaborators); (on the output side) researchers have different inclinations to codify their results under forms other than publication, or there are divergent agglomeration[10] or scope economies.

## 4. Survey model

The surveyed field includes all the Italian universities with at least 4 employed resources during the surveyed years (including full professors, associate professors and research fellows)[11] in scientific-technological University Disciplinary Areas (UDA)[12], for which the SCI database can be extensively used. The decoding of surveyed UDAs is shown in Table 1. The study period includes years 2001 to 2003.

[Table 1]

The bibliometric non-parametric methodology we adopted involves application of the DEA technique to bibliometric data regarding the scientific production of all national universities for each UDA of activity. The reason for choosing the DEA technique came from what was observed in sections 2 ad 3 of this paper regarding the comparative advantages of bibliometric methodologies over peer-review ones (for instance, objective metrics and results, simpler application (including intertemporal application), non-invasiveness and low costs), and within bibliometric methodologies, its comparative advantages over non-parametric techniques (no assumptions on the production function, possible identification of best performers among the DMUs under study, possible calculation of total factor productivity indexes). The DEA technique also eliminates any distortion in the productivity measurement due to possible variable returns to scale of the production factors.

---

[10] A host of studies have demonstrated the positive effect of proximity of private research on the research productivity of public laboratories (Siegel et al., 2003).

[11] The definition of such threshold was made necessary by the empirical observation that the values of total factor productivity might be distorted by input values being all close to zero.

[12] The Italian university system adopts a classification system comprising 14 "areas", which in turn include 370 "disciplinary sectors". See for details http://www.miur.it/UserFiles/115.htm. Of the whole, we have surveyed the 9 science areas, including 205 scientific-disciplinary sectors. It should be noted that the "Civil Engineering and Architecture" area straddles the border between scientific-technological disciplines (typical of civil engineering) and art disciplines (typical of some sectors of architecture).

The input and output variables used to feed the DEA model, obtained as the average of the point values from the 2001-2003 period, are shown in Table 2 and described in what follows.

[Table 2]

- The FP, AP and RF inputs were obtained from the list of the scientific staff employed at Italian universities, which is maintained by the Italian Department for Research and University Education. The choice of separating the various types of internal staff had the purpose of distinguishing different degrees of "quality" among the employed human resources. A significant productivity differential by function has accordingly been shown in the literature (Prpic, 1996; Zainab, 1999; Bordons et al., 2003).
- The PR input includes additional financial resources for research, which are potential determinants of increased scientific production.[13] For that variable, a proxy measure was used, i.e. the funds for the National Interest Research Projects (PRIN – http://prin.miur.it)[14], a program intended for universities alone and managed centrally by the Ministry for Research and University Education, which grants support on areas acknowledged as strategic for the development of the country. Within the University accounting systems, PRIN funds are the only ones that can be unambiguously matched with each single DMU. They account for about 15% of total additional funds for universities. The authors have revealed a significant correlation between PRIN and the total university-level additional funds, and are therefore confident that such significance could be assumed at the level of individual disciplinary areas.

---

[13] Ordinary Funds, unlike additional resources, are usually aimed at ensuring operation of the university and covering labor costs; it is therefore reasonable to assume homogeneous pro-capita resources for each university. Correlated input vectors (human-resource-correlated, in this case) will not alter the result of DEA calculations, and may therefore be excluded from the model.

[14] For the period under exam, the authors have verified a significant correlation between PRIN funds and total additional funds at university level, and are therefore confident that such significance can be assumed at the level of individual disciplinary areas. In fact, no other data on the allocation of financial resources by universities bears a relationship with individual disciplinary areas.

- The PU output, for the *i*-th UDA of the *j*-th university, is calculated as the sum of publications with at least one author from University *j* belonging to Area *i*.
- The PC output is a similar index to the PU, but takes into account authors' "contribution", measured as the ratio between the number of authors belonging to that UDA and the total number of authors of the publication:

$$PC_{ij} = \sum_{pubblicazioni} \frac{b_{ij}}{c}$$

where $b_{ij}$ equals the sum of the numbers of authors of the publication belonging to the *i*-th UDA of the *j*-th University, and *c* is the total number of authors of the publication.

- The SS output (scientific strenght) equals the weighted average of total publications by each university within each UDA. The weights, in particular, are referred to the impact factor of the journal in which each publication is included.[15]

The input values are recorded as of 31 December of the year prior to output record.

Data on output variables were extracted by the ORP (Osservatorio sulla Ricerca Pubblica - Observatory on Public Research) database, developed to this purpose by the authors, from Thomson Scientific's SCI database. It collects and sorts out information of the scientific production by researchers from Italian universities, while enabling aggregation operations at higher levels (UDA, School, University), with better degrees of accuracy than those obtained with aggregated extractions by university.

By processing and post-coding of SCI data, the observatory surveys the scientific production (papers and reviews only) by all 77 Italian universities during the 2001-2003 period. The survey is performed by name, as it refers to publications by scientific staff (full professors, associate professors and research scientists)[16] in the universities. It was therefore necessary to homogenize all possible designations for Italian universities. As mentioned in section 2.2, one of the reasons for errors in publication attribution is that different denominations are used by authors to identify the same institution. Such homogenization made it possible to extract from the SCI$^{TM}$ database the universe of publications (62,523 in the 2001-2003 period) with at least one "address" compatible with those of Italian universities. Subsequently, the designations included in the author list of those publications were disambiguated: to that end, a specific rule-based algorithm was formulated and implemented, which helped retrace the publication-author-university-UDA link by coupling the author list of the above-mentioned 62,523 publications with the database of scientific university staff. Such procedure proved particularly taxing,[17] first because the records in the SCI$^{TM}$ show no link between the "author list" and the "address list" of a publication, and secondly because of particularly strong homonymy, which, in turn, results from two distinct factors: the coding of names in the "author list" of SCI$^{TM}$ (last name and initial of first name), and the size of the observed population (over 36,000 university research staff). It should be noted that about half of the 62,523 publications originally surveyed showed homonymy cases in its author list. With the developed algorithm, it was possible to disambiguate 53,420 publications; for 1,593, manual disambiguation was necessary. The remaining 9,298 publications, among those originally surveyed, were discarded. Those were papers by authors who were not included in the list of university research staff but had stated affiliation to a university (9,103), and a small minority of publications (195) which could not be disambiguated by means of the algorithm.

---

[15] This indicator represents a proxy measure of the total number of citations traceable to the scientific production of the unit under consideration.

[16] While other research profiles exist in universities, such as non-tenure teaching staff, graduate students,

In this study, the aggregation of output data by disciplinary area is performed by simply adding the data pertaining to each author within the specific area she/he belongs to. This is a novel approach in this area of study, especially considering how broad and diverse the study field is.

Table 3 shows the descriptive statistics of the analyzed input and output values, classified by UDA.

---

undergraduates etc., these are not identifiable.

[17] To the authors' knowledge, only two studies on the disambiguation of scientific publications were published: one of them (Wooding et al., 2006) utilizes a recursive algorithm for disambiguating publications by researchers of arthritic diseases; the other (Torvik et al., 2005) employs a stochastic similarity measure applied to the publications contained in the Medline© repertoire of the American National Library of Medicine.

[Table 3]

**5. Results of the analysis and observations**

The DEA was applied to the input and output data obtained as described above for each university and, within each university, for each UDA, under the hypotheses of: i) output orientation; ii) radial measure of efficiency. Table 4 presents the statistics pertaining to Technical Efficiency, Pure Technical Efficiency and Scale Efficiency identified for every university operating within each surveyed area. The unit value is typical of efficient units. In general, the score assigned to each DMU is equal to the reciprocal of the equal proportional increase of all outputs (inputs remaining equal) needed to reach a certain position on the frontier.

Overall, differences in average efficiency and score variability (measured in terms of standard deviation) among different UDAs are especially significant. Such differences are clearly a sign of varying scientific prolificity among UDAs; yet, a possible distorting effect should be taken into account, resulting from the different number of journals surveyed among UDAs and the representativeness of the SCI as a reference universe for the surveying of scientific production in all areas.[18] Returns to scale also show erratic distributions among different UDAs, evidenced by a highly variable number of universities with constant returns to scale. Similarly, the number of universities with variable returns to scale has an erratic distribution among UDAs.

[Table 4]

Table 5 presents the data from the group analysis based on PTE scores, distributed by UDA as usual. The performance difference between university groups is rather evident, with average score variations of 17%, 35% and 56% between efficient universities and respectively, the first, second and third percentile of inefficient ones. In this calculation, UDA 8 was excluded as it presented rather anomalous technical efficiency data. Such anomaly might be explained considering what was said earlier about the representativeness of the SCI database for this particular UDA. Preceding observations confirm our preliminary warning against performing aggregated calculations at university level rather than at individual disciplinary area level, if the inconsistency intrinsic to the DMUs of the model is to be taken into consideration.

---

[18] While journals in so-called border disciplines, such as those included in Management Engineering, Architecture etc., reach similar quality standards, they might happen not be included in the SCI, but rather in other Thomson databanks such as the Social Science Citation Index or the Art and Humanities Citation Index.

[Table 5]

Based on the efficiency indexes by UDA assigned to the universities, a global efficiency index was calculated for each university, as a weighted index of scores at UDA level, normalized to their respective mean values.[19] The weighting was performed on the basis of the weight of each UDA within each university, evaluated in terms of employed research personnel. Let $\theta_{ij}$ be the efficiency score (PTE) of the i-th university in the j-th UDA, normalized to the average efficiency in the j-th UDA, and let $R_{ij}$ be the total number of employed research staff (sum of FP, AP and RF); the total score for the i-th university is thus obtained as:

$$\Theta_{Tot}(i) = \frac{\sum_{ADU=j} \Theta_{ij} R_{ij}}{\sum_{ADU=j} R_{ij}}$$

Table 6 shows the university ranking, both general and by active UDA. The top positions are always occupied by the more efficient universities, which lie at the frontier.

The strong variability in the ranking of some universities in different UDAs might be due to production factors not included in the model affecting the productivity in each UDA in different ways, and to localization or scope economies. The residual value, whose size is not calculable, can be explained as a result of the different quality of human resources, and would seem to denote the degree of management integration of the whole organization: in fact, a vision of excellence is presumably not compatible with a strong variability in human resource quality among UDAs in the same university.

---

[19] Such normalization is aimed at offsetting the potential distorting effect of the strong variability in average efficiency values among different Disciplinary Areas.

[Table 6]

The results of the application of the proposed model were compared to those deriving from a simplified single input/single output type model, in which productivity is calculated as the ratio between the number of publications and the total number of employees. The data on ranking variation between the two methods are presented in Table 7, and show remarkable variation values for all UDAs; that indicates that both the complication of the model by clarification of input resource types (researchers, associate professors and full professors and output resource types (total, contribution-based, qualitative), and the presence of variable returns to scale can alter partial productivity rankings significantly.

[Table 7]

In order to assess the sensitivity of the model with changing independent variables, an analysis of ranking variations resulting from the exclusion of the "additional funds" variable from the model inputs was also conducted (Table 8). The comparison shows strong ranking variability especially in UDAs 1, 5, 6 and 9. The number of efficient DMU decreases, thus determining an "impoverishment" of the frontier. The same is not true for the sensitivity to the number of outputs taken into consideration, as there is a strong correlation among output values, i.e. the assessment dimensions (qualitative, quantitative and contribution-based) of scientific output.

[Table 8]

## 6. Conclusions

This study was aimed at developing and validating a bibliometric non-parametric methodology for measuring the performance of public research activity. In the context of the international state of the art, the most severe limitations of bibliometric approaches were overcome thanks to significant improvements: i) the DEA model that was developed takes into consideration the incidence of the different forms of input and output in the research activity; in particular, among outputs all values used in the measurement of scientific production (quantity, quality and contribution) were considered; ii) the assessment of efficiency at individual disciplinary area level limits distortions due to heterogeneity in terms of resource mixes present in the universities and different peculiarities of each area (prolificity and representativeness in source databases); most importantly, iii) the adoption of a "bottom-up" approach, in which output values are associated to each individual author and are subsequently aggregated by UDA-University, with a structured homogenization and disambiguation procedure, ensures accuracy levels never attained before in large-scale studies in the literature. The results obtained show strong heterogeneity in average performances among different UDAs and evidence, for each university, the main dimension of inefficiency (scale, SE, or resource use, PTE) and consequent management remedial measures.

This study does not pretend to be exhaustive or definitive. Our analyses present strong sensitivity of the results to the type of values used as inputs and outputs of the model, which requires further methodological investigation and fine-tuning; the authors are currently working on that issue. Furthermore, the sensitivity of the model to the values used imposes caution, especially as the possible implications of using such methodology to support the resource allocation procedure are taken into consideration.

Regardless of the degree of reliability of different assessment models, compared analyses of research productivity in universities and, more generally, public research institutions, deserve careful study, and even require it in case they are liable to be used by policy makers for allocation purposes. A first question to ask is whether scarce resources should be allocated to public universities according a more or less invariant excellence concept, as implied by the definition of universal and undistinguished algorithms and methods of measurement, or rather according a more articulated set of strategic criteria, which vary over space and time. The heterogeneity in location, culture, size and specialization among universities would seem to require the strategies developed by individual universities to be necessarily differentiated, and in some cases potentially complementary. The strategic perspective in resource allocation entails that universities might have different strategic objectives, and should therefore not necessarily be assessed in a uniform manner. It is also important to acknowledge that universities present a unique organizational specificity: their members operate at the same time in two areas, namely teaching and research. Depending on the distinctive competences of each university, and the typical needs of the area where it is located, it might be appropriate to differentiate the emphasis placed on different activities or within each of them, to pursue different objectives with varying strength. Within a single type of activity, such as research, one disciplinary area could be favoured over another, irrespective of the knowledge level in that area; opposite situations might reveal appropriate in different areas of the country.

In spite of all that, outcome control assessments are extensively used to measure university performances and, in some cases, to influence allocation decisions by policy makers. As it overcomes the representativeness limitations described above, our model therefore forms a robust benchmarking tool for research management and a valid integrating instrument to support policy makers' decisions.

Further analyses and developments of the proposed methodology, part of which have already been undertaken by the authors, may concern: (in terms of methodological fine-tuning) i) use of production factors cost vectors to estimate economic and allocation efficiency; ii) use of citations rather than the impact factor in assessing the quality of a paper; iii) integration of data from the 2004-2006 period for time series analyses and extension to the socio-economic and art and humanities areas. It would also be possible (and welcome) to analyze lower aggregation levels, for instance individual scientific-disciplinary sectors. Finally, the "determinants" of empirically identified performances could be studied more specifically: in particular, an interesting development would be detecting economies of scope or localization impact, and analyzing possible differentials by function, gender and general personal characteristics.

| DISCIPLINARY AREA | DESIGNATION |
|---|---|
| UDA 1 | Mathematical sciences |
| UDA 2 | Physical sciences |
| UDA 3 | Chemical sciences |
| UDA 4 | Earth sciences |
| UDA 5 | Biological sciences |
| UDA 6 | Medical sciences |
| UDA 7 | Agricultural and veterinary sciences |
| UDA 8 | Civil engineering and architecture |
| UDA 9 | Industrial and information engineering |

*Table 1: Science University Disciplinary Areas.*

| VARIABLE | TYPE | ACRONYM |
|---|---|---|
| Number of full professors | Input | FP |
| Number of associate professors | Input | AP |
| Number of research scientists | Input | RF |
| PRIN funding for research (k€) | Input | PR |
| Number of publications | Output | PU |
| Contribution to publications | Output | PC |
| Scientific strength | Output | SS |

*Table 2: Variable of the DEA survey model used.*

|     | UDA1 (52 Universities) | | | | UDA2 (49 Universities) | | | | UDA3 (47 Universities) | | | |
| --- | --- | --- | --- | --- | --- | --- | --- | --- | --- | --- | --- | --- |
|     | Ave | Min | Max | Std.Dev | Ave | Min | Max | Std.Dev | Ave | Min | Max | Std.Dev |
| FP | 18 | 0 | 86 | 18 | 16 | 1 | 72 | 15 | 20 | 0 | 69 | 19 |
| AP | 21 | 1 | 109 | 20 | 19 | 0 | 62 | 17 | 25 | 1 | 103 | 22 |
| RF | 21 | 2 | 69 | 17 | 16 | 1 | 52 | 12 | 22 | 2 | 81 | 19 |
| PR | 90.86 | 0 | 374.6 | 99.7 | 199.1 | 0 | 948.21 | 182.67 | 303 | 0 | 1.148.15 | 299.35 |
| PU | 21 | 1 | 94 | 19 | 56 | 1 | 260 | 50 | 87 | 4 | 367 | 79 |
| PC | 13 | 1 | 55 | 11 | 19 | 0 | 86 | 17 | 49 | 2 | 226 | 46 |
| SS | 24 | 0 | 110 | 24 | 150 | 4 | 724 | 147 | 220 | 10 | 995 | 213 |
|     | UDA4 (39 Universities) | | | | UDA5 (53 Universities) | | | | UDA6 (43 Universities) | | | |
|     | Ave | Min | Max | Std.Dev | Ave | Min | Max | Std.Dev | Ave | Min | Max | Std.Dev |
| FP | 10 | 1 | 33 | 8 | 27 | 0 | 103 | 24 | 57 | 0 | 226 | 45 |
| AP | 12 | 1 | 37 | 9 | 30 | 0 | 117 | 28 | 77 | 1 | 385 | 72 |
| RF | 11 | 1 | 32 | 7 | 35 | 0 | 154 | 31 | 111 | 1 | 688 | 128 |
| PR | 94.59 | 0 | 350.5 | 85.28 | 302.3 | 0 | 1.407.20 | 314.09 | 557.8 | 15.7 | 1.925.21 | 463.67 |
| PU | 15 | 0 | 41 | 12 | 79 | 2 | 317 | 69 | 166 | 2 | 653 | 144 |
| PC | 7 | 0 | 22 | 6 | 42 | 1 | 175 | 38 | 87 | 1 | 328 | 76 |
| SS | 29 | 0 | 101 | 25 | 275 | 5 | 1120 | 254 | 578 | 5 | 2230 | 507 |
|     | UDA7 (33 Universities) | | | | UDA8 (37 Universities) | | | | UDA9 (46 Universities) | | | |
|     | Ave | Min | Max | Std.Dev | Ave | Min | Max | Std.Dev | Ave | Min | Max | Std.Dev |
| FP | 29 | 1 | 86 | 26 | 27 | 1 | 135 | 31 | 33 | 1 | 169 | 40 |
| AP | 27 | 0 | 80 | 23 | 32 | 2 | 134 | 33 | 31 | 1 | 161 | 36 |
| RF | 34 | 1 | 112 | 30 | 36 | 2 | 149 | 38 | 29 | 0 | 136 | 30 |
| PR | 228.3 | 0 | 915.8 | 220.94 | 225.7 | 4.1 | 1156.51 | 226.17 | 328.5 | 0 | 1.510.28 | 366.84 |
| PU | 30 | 0 | 116 | 30 | 11 | 1 | 45 | 10 | 45 | 0 | 203 | 45 |
| PC | 17 | 0 | 76 | 19 | 7 | 1 | 29 | 6 | 28 | 0 | 131 | 29 |
| SS | 56 | 0 | 226 | 60 | 16 | 1 | 55 | 15 | 66 | 0 | 295 | 68 |

*Table 3: Descriptive statistics of model input and output values (average values during 2001-2003 period).*

|  | UDA | 1 | 2 | 3 | 4 | 5 | 6 | 7 | 8 | 9 |
|---|---|---|---|---|---|---|---|---|---|---|
|  | No. of universities | 52 | 49 | 47 | 39 | 53 | 43 | 33 | 37 | 46 |
|  | Efficient universities | 15 | 8 | 11 | 5 | 12 | 16 | 10 | 1 | 13 |
| TE | *Ave* | 0.719 | 0.668 | 0.793 | 0.639 | 0.737 | 0.860 | 0.752 | 0.334 | 0.710 |
|  | *Min* | 0.353 | 0.154 | 0.406 | 0.123 | 0.343 | 0.314 | 0.259 | 0.017 | 0.263 |
|  | *Std.Dev* | 0.222 | 0.230 | 0.172 | 0.230 | 0.190 | 0.169 | 0.242 | 0.237 | 0.236 |
| PTE | *Ave* | 0.855 | 0.827 | 0.883 | 0.793 | 0.871 | 0.897 | 0.852 | 0.660 | 0.843 |
|  | *Min* | 0.405 | 0.405 | 0.449 | 0.198 | 0.369 | 0.385 | 0.370 | 0.094 | 0.303 |
|  | *Std.Dev* | 0.184 | 0.202 | 0.148 | 0.242 | 0.160 | 0.153 | 0.195 | 0.285 | 0.192 |
| SE | *Ave* | 0.840 | 0.804 | 0.896 | 0.813 | 0.844 | 0.959 | 0.866 | 0.492 | 0.838 |
|  | *Min* | 0.484 | 0.274 | 0.571 | 0.338 | 0.528 | 0.451 | 0.636 | 0.139 | 0.470 |
|  | *Std.Dev* | 0.158 | 0.172 | 0.098 | 0.169 | 0.129 | 0.097 | 0.122 | 0.231 | 0.171 |
| *CRS* |  | 15 | 8 | 11 | 6 | 12 | 16 | 10 | 2 | 13 |
| *NIR* |  | 31 | 33 | 25 | 27 | 38 | 12 | 21 | 34 | 29 |
| *NDR* |  | 6 | 8 | 11 | 6 | 3 | 15 | 2 | 1 | 4 |

*Table 4: Statistics of Technical Efficiency, Pure Technical Efficiency and Scale Efficiency by UDA.*

| DA | Efficient universities | Inefficient universities | | |
|---|---|---|---|---|
|  |  | 1st tertile | 2nd tertile | 3rd tertile |
| 1 | 25 (out of 52) | 0.888 | 0.744 | 0.529 |
| 2 | 19 (out of 49) | 0.926 | 0.735 | 0.493 |
| 3 | 18 (out of 47) | 0.965 | 0.830 | 0.651 |
| 4 | 15 (out of 39) | 0.905 | 0.684 | 0.404 |
| 5 | 23 (out of 53) | 0.930 | 0.770 | 0.616 |
| 6 | 21 (out of 43) | 0.948 | 0.840 | 0.633 |
| 7 | 17 (out of 33) | 0.854 | 0.756 | 0.513 |
| 8 | 9 (out of 37) | 0.813 | 0.582 | 0.291 |
| 9 | 19 (out of 46) | 0.913 | 0.763 | 0.522 |

*Table 5: Average values of Pure Technical Efficiency by university groups.*

| University code | General ranking | Ranking for individual UDAs | | | | | | | | |
|---|---|---|---|---|---|---|---|---|---|---|
| | | 1 | 2 | 3 | 4 | 5 | 6 | 7 | 8 | 9 |
| 1 | 1 | - | 1 | - | 1 | 1 | - | - | - | 1 |
| 2 | 1 | - | - | - | - | 1 | - | - | - | - |
| 3 | 1 | 1 | 1 | - | - | 1 | - | - | - | - |
| 4 | 1 | - | - | - | - | - | - | 1 | - | 1 |
| 5 | 1 | - | - | - | - | - | - | - | - | 1 |
| 6 | 1 | - | - | - | - | 1 | - | 1 | - | - |
| 7 | 1 | 1 | 1 | 1 | - | - | - | 1 | 1 | 1 |
| 8 | 1 | - | - | - | - | 1 | 1 | - | - | - |
| 9 | 9 | 1 | 22 | 23 | 1 | 1 | 1 | 1 | - | 1 |
| 10 | 10 | 1 | 1 | 29 | 1 | 1 | 1 | 1 | 1 | 20 |
| 11 | 11 | 41 | 1 | - | - | 1 | 1 | 1 | - | 1 |
| 12 | 12 | 1 | 21 | 1 | 1 | 24 | 1 | 1 | 19 | 25 |
| 13 | 13 | 48 | 38 | - | - | 1 | 1 | 22 | - | - |
| 14 | 14 | 29 | 27 | 27 | 1 | 1 | 1 | 19 | - | - |
| 15 | 15 | - | - | - | - | - | 25 | - | - | - |
| 16 | 16 | 1 | 28 | - | - | 1 | - | - | - | - |
| 17 | 17 | 1 | 31 | 1 | - | 1 | 23 | - | 29 | 1 |
| 18 | 18 | 32 | 1 | 37 | 22 | 1 | 1 | 24 | 1 | 1 |
| 19 | 19 | 1 | 33 | 1 | 17 | 33 | 1 | - | - | - |
| 20 | 20 | 28 | 1 | 1 | 32 | - | - | - | 1 | 26 |
| 21 | 21 | 46 | 37 | 45 | - | - | - | - | 1 | 1 |
| 22 | 22 | 1 | 46 | 1 | 1 | 30 | 26 | 1 | - | 1 |
| 23 | 23 | 53 | 1 | 36 | 25 | 1 | 1 | 1 | 11 | 1 |
| 24 | 24 | 1 | 1 | 1 | - | 43 | 1 | - | - | - |
| 25 | 25 | 38 | 35 | 1 | 1 | 31 | 1 | 1 | 17 | 23 |
| 26 | 26 | 39 | 20 | 1 | 1 | 1 | 27 | 23 | 16 | 33 |
| 27 | 27 | - | - | 1 | - | 1 | 34 | 1 | - | - |
| 28 | 28 | 34 | 23 | 32 | 16 | 27 | 1 | 25 | 27 | 29 |
| 29 | 29 | 36 | 25 | 43 | 1 | 41 | 1 | - | 1 | 27 |
| 30 | 30 | 52 | 34 | 1 | 23 | 1 | 1 | - | 23 | 1 |
| 31 | 31 | 1 | 42 | 25 | 21 | 44 | 24 | - | 20 | 1 |
| 32 | 32 | 1 | 1 | 1 | - | 35 | 30 | - | 28 | 36 |
| 33 | 33 | 1 | 1 | 47 | 1 | 1 | - | - | 21 | 21 |
| 34 | 34 | 1 | 1 | 35 | 1 | 45 | 37 | - | 25 | 1 |
| 35 | 35 | 1 | 49 | 21 | - | 37 | 29 | - | - | - |
| 36 | 36 | 44 | 36 | 1 | 33 | 1 | 1 | - | 18 | 24 |
| 37 | 37 | - | - | 19 | - | 48 | - | 18 | - | - |
| 38 | 38 | 35 | 44 | 1 | 19 | 34 | 31 | 29 | 24 | 1 |
| 39 | 39 | 47 | 1 | 28 | 30 | 29 | 39 | 1 | 1 | 31 |
| 40 | 40 | 42 | 47 | 26 | 36 | 36 | 1 | 20 | 22 | 28 |
| 41 | 41 | 1 | 39 | 46 | 28 | 28 | 32 | 26 | - | 1 |
| 42 | 42 | 31 | 1 | 1 | - | 1 | - | 1 | 30 | 38 |
| 43 | 43 | - | - | 1 | - | 47 | 35 | - | - | - |
| 44 | 44 | 51 | 26 | 30 | 20 | 40 | 40 | 21 | 15 | 30 |
| 45 | 45 | 27 | 30 | 39 | 29 | 50 | 28 | 31 | 26 | 22 |
| 46 | 46 | 40 | 45 | 40 | 27 | 1 | 38 | - | 1 | 37 |
| 47 | 47 | 45 | 32 | 24 | 35 | 42 | 33 | - | 31 | 32 |
| 48 | 48 | 30 | 24 | 33 | 24 | 49 | 1 | 33 | 14 | 42 |
| 49 | 49 | 49 | 1 | 42 | 1 | 1 | - | - | - | 1 |

| University code | General ranking | Ranking for individual UDAs | | | | | | | | |
|---|---|---|---|---|---|---|---|---|---|---|
| | | 1 | 2 | 3 | 4 | 5 | 6 | 7 | 8 | 9 |
| 50 | 50 | - | - | 1 | 1 | 53 | - | 1 | - | - |
| 51 | 51 | 1 | - | - | - | - | 1 | 1 | 12 | 39 |
| 52 | 52 | 1 | - | - | 26 | 39 | - | - | - | 41 |
| 53 | 53 | 1 | 43 | 44 | 34 | 38 | 41 | 1 | 1 | 35 |
| 54 | 54 | 1 | 41 | - | - | 32 | 1 | - | 13 | 46 |
| 55 | 55 | 50 | 29 | 34 | 31 | 52 | 36 | - | - | 1 |
| 56 | 56 | - | 1 | 38 | 1 | 51 | 42 | 28 | - | - |
| 57 | 57 | 33 | 1 | 1 | 1 | 26 | 1 | 30 | 32 | 45 |
| 58 | 58 | 1 | 1 | 20 | 37 | 25 | - | 27 | 35 | 43 |
| 59 | 59 | 1 | - | - | - | - | - | - | - | 44 |
| 60 | 60 | 1 | 40 | 31 | 18 | 46 | 43 | 1 | 38 | - |
| 61 | 61 | 43 | 1 | 41 | 38 | 1 | 22 | - | 36 | - |
| 62 | 62 | 37 | 48 | 22 | 39 | - | - | - | 33 | 40 |
| 63 | 63 | 1 | - | - | - | - | - | 32 | 34 | 34 |
| 64 | 64 | 1 | - | - | - | - | - | - | 37 | 1 |

*Table 6: General ranking and ranking by individual UDAs (for each University) in terms of Pure Technical Efficiency.*

| UDA | Variations | Max variation | Average variation | Median | Var. coeff. |
|---|---|---|---|---|---|
| 1 | 51 (out of 52) | 51 | 14 | 10 | 0.94 |
| 2 | 47 (out of 49) | 44 | 10 | 7 | 0.99 |
| 3 | 43 (out of 47) | 44 | 10 | 9 | 0.94 |
| 4 | 37 (out of 39) | 36 | 9 | 5 | 1.06 |
| 5 | 50 (out of 53) | 50 | 13 | 8 | 1.02 |
| 6 | 41 (out of 43) | 42 | 13 | 10 | 0.86 |
| 7 | 32 (out of 33) | 32 | 8 | 5 | 1.09 |
| 8 | 34 (out of 37) | 26 | 8 | 7 | 0.95 |
| 9 | 44 (out of 46) | 45 | 12 | 10 | 1.01 |

*Table 7: Compared rankings of Pure Technical Efficiency and single output/single input (number of publications/total number of research staff).*

| UDA | Variations | No longer efficient universities | Max variation | Average variation | Median | Var. coeff. |
|---|---|---|---|---|---|---|
| 1 | 46 (out of 52) | 7 | 42 | 6 | 5 | 1.07 |
| 2 | 37 (out of 49) | 6 | 22 | 3 | 2 | 1.13 |
| 3 | 40 (out of 47) | 1 | 12 | 2 | 2 | 1.01 |
| 4 | 19 (out of 39) | 1 | 5 | 1 | 0 | 1.45 |
| 5 | 43 (out of 53) | 6 | 42 | 5 | 4 | 1.39 |
| 6 | 38 (out of 43) | 8 | 33 | 6 | 4 | 1.14 |
| 7 | 25 (out of 33) | 2 | 10 | 2 | 1 | 1.31 |
| 8 | 25 (out of 37) | 1 | 6 | 2 | 1 | 1.02 |
| 9 | 42 (out of 46) | 3 | 45 | 7 | 5 | 1.47 |

*Table 8: Analysis of sensitivity to the presence of an "additional funding" variable among the DEA model inputs. Ranking variations in terms of Pure Technical Efficiency scores.*